\documentclass[conference, 12pt]{IEEEtran}
\usepackage{setspace}
\usepackage{amsmath,amsfonts}
\usepackage{algorithmic}
\usepackage{algorithm}
\usepackage{array}
\usepackage[caption=false,font=normalsize,labelfont=sf,textfont=sf]{subfig}
\usepackage{textcomp}
\usepackage{stfloats}
\usepackage{url}
\usepackage{verbatim}
\usepackage{graphicx}
\usepackage{cite}
\usepackage{hyperref}
\usepackage{cleveref}
\usepackage{orcidlink}
\usepackage{microtype}
\begin{document}
\onecolumn
\title{DRL-Based Secure Spectrum-Reuse D2D Communications with RIS Assistance}

\author{Maryam Asadi Ahmadabadi \orcidlink{0009-0001-8100-8316}, Farimehr Zohari \orcidlink{0000-0003-2039-8135}, S. Mohammad Razavizadeh \orcidlink{0000-0002-1732-4779}~\IEEEmembership{Senior Member,~IEEE,}}
\maketitle
\begin{abstract}
This study examines the secrecy performance of an uplink device-to-device (D2D) communication system enhanced by reconfigurable intelligent surfaces (RIS) while considering the presence of multiple eavesdroppers. RIS technology is employed to improve wireless communication environment by intelligently reflecting signals, thereby improving both capacity and security. We employ deep reinforcement learning (DRL) to optimize resource allocation dynamically, addressing challenges in D2D pairs and optimizing RIS positioning and phase shifts in a changing wireless environment. Our simulations demonstrate that the developed DRL-based framework significantly maximizes the sum secrecy capacity of both D2D and cellular communications, achieving higher transmission secrecy rates compared to existing benchmarks. The results highlight the effectiveness of integrating RIS with D2D communications for improved security performance.
\end{abstract}


\section{Introduction}
\IEEEPARstart{D}{evice}-to-device (D2D) communication, also known as proximity services (ProSe), plays an important role in be yond 5G networks to enhance the overall user experience and network performance \cite{1}. Despite the numerous advantages of D2D communications, they have some challenges that need to be addressed for successful implementation. One of the key challenges is ensuring the security of data communications in D2D networks, as an eavesdropper might easily intercept direct D2D interactions. To address this challenge and strengthen the security of D2D communications, physical layer security (PLS) Has evolved into a promising approach that can be integrated with higher-layer security mechanisms. Several studies have explored the application of PLS in the context of D2D communications, e.g. \cite{2,3,4}. The introduction of reconfigurable intelligent surfaces (RIS) represents a significant breakthrough in 6G wireless networks, as they have great potential to improve data rates, optimize spectrum usage, and improve energy efficiency \cite{5}. RISs can also be utilized to enhance PLS in wireless communication systems and increase the system’s resistance to eavesdropping and intentional interference. Numerous works have investigated RIS-assisted PLS in centralized wireless networks, e.g. \cite{6,7,8,9}. Although extensive research has explored RIS-assisted PLS in centralized wireless networks, only a limited number of studies have addressed its application in D2D wireless networks. For example, in \cite{13}, RIS technology is investigated to enhance the reliability and robustness of D2D communication and to improve the level of security of the cellular network simultaneously. New analytical expressions are derived for the cellular secrecy outage probability (SOP), the D2D outage probability, and the probability of nonzero secrecy capacity (PNSC). Further contributions include \cite{15} that evaluates the secrecy performance of an RIS-D2D communication in cellular networks shared by the spectrum. To achieve this, new closed-form expressions for the SOP and the asymptotic SOP are derived considering the presence of multiple eavesdroppers. Moreover, \cite{14} explores PLS and data transmission in underlay D2D networks, examining a combination of RIS and full-duplex jamming receivers to improve the system's robustness and security. The work in \cite{10726861} examined the performance of PLS in dual RIS-aided V2V NOMA communication systems. Lastly, \cite{10683052} investigates the energy-constrained D2D transmitter in cognitive cellular networks for secure D2D communication to a targeted D2D receiver.\\ 
The studies presented in the aforementioned works cover various aspects within the security of RIS-assisted D2D communication in cellular networks. In existing studies, no comprehensive system model that simultaneously addresses secure performance for both D2D and cellular users has been adequately explored in a network with multiple cellular and D2D users. Additionally, determining the optimal location for deploying the RIS to enhance the system's secure performance remains a critical challenge. To bridge these gaps, we investigate PLS in an uplink RIS-aided underlay D2D communication system. Specifically, our focus is on ensuring secure data transmission while optimizing the RIS placement to maximize the system's sum secrecy capacity (SSC) in a dynamic environment. To the best of our knowledge, this issue has not been extensively explored in existing literature.\\
Our approach involves formulating an optimization problem under constraints, including secure rate requirements for both cellular and D2D users, the transmission power of D2D transmitters, the RIS phase-shift matrix, and the RIS location. This problem is particularly challenging due to its non-convex nature, stemming from resource allocation restrictions among D2D pairs. The nonconvexity is further exacerbated by the dynamic environment, making conventional mathematical optimization techniques impractical. Methods such as Semidefinite Relaxation (SDR), Minorization-Maximization (MM), and Block Coordinate Descent (BCD) struggle to adapt to rapid variations in channel conditions and uncertainties in channel state information (CSI). Additionally, their computational complexity increases significantly with the number of users and RIS elements, which limits their scalability. Consequently, meeting the exacting requirements of emerging 6G network applications becomes an increasingly formidable challenge.\\
To address this non-convex problem, one potential solution lies in using artificial intelligence (AI) and machine learning (ML) techniques. These methods can adaptively learn and optimize network parameters based on observed data, allowing for dynamic adjustments to changing channel conditions and uncertainties in CSI. Among ML-based methods, deep reinforcement learning (DRL) has been proposed in the literature for optimizing network parameters and improving the performance of RIS-assisted networks \cite{9729826}. In this work, we use DRL algorithms to solve our challenging problem. Specifically, considering hardware limitations and a close-to-practical model, we assume that RIS elements can only provide discrete phase shifts. Thus, our problem is investigated in the discrete domain. To achieve this, we employ a double deep Q-network (DDQN) technique due to its stability and faster learning speed. Furthermore, to mitigate the high computational burden of centralized intermittent optimization algorithms on the central node, a distributed optimization approach using decentralized DDQN is adopted for D2D resource allocation. This enables efficient and scalable resource allocation across the network. Finally, the effectiveness of this approach is demonstrated on the proposed system model. Our results verify that this method has a significant performance advantage over other approaches, such as the deep Q-network (DQN) method, due to its ability to address the overestimation problem.
\begin{figure}
    \centering
    \includegraphics[width=0.7\linewidth]{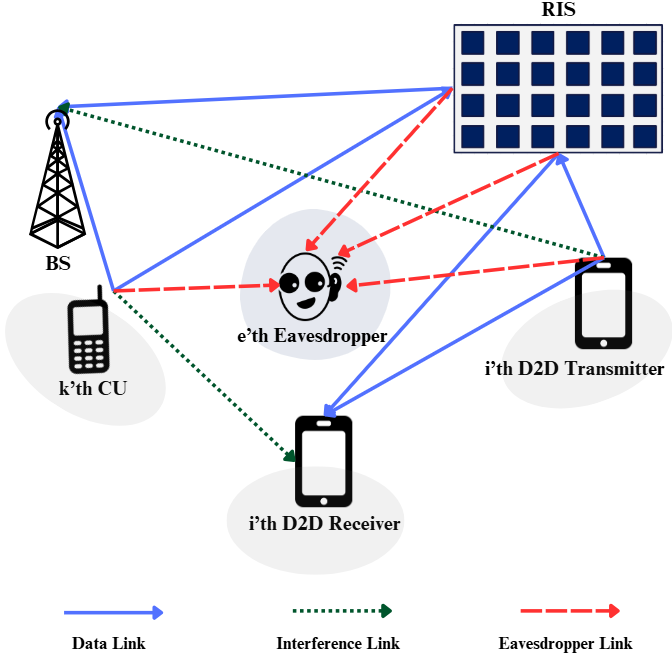}
    \caption{RIS-aided D2D communication in cellular networks in the presence of eavesdroppers.}
    \label{fig:enter-label}
\end{figure}
\section{System Model and Problem Formulation}
As illustrated in Fig. 1, we consider a D2D-underlaid uplink cellular system consisting a BS, $K$ cellular users (CU), $M$ D2D pairs and $E$ eavesdroppers. Each CU occupies one subchannel and in order to improve spectrum efficiency, the subchannel of the CU can be reused by a pair of D2D. we assume the eavesdroppers to be non-collusive. To optimize the communication environment and resource allocation for D2D pairs simultaneously, we use a RIS that comprises N passive elements arranged in a uniform linear array (ULA).\\
The channel structure consists of two components: a direct link and a reflective link generated by the RIS.
When the $i$'th D2D transmitter ($D_i^t$) establishes communication with its corresponding receiver ($D_i^r$) by utilizing the resource block (RB) of the $k$'th cellular user, the channel coefficient denoted as
\begin{equation}
\begin{array}{l}
h_i[k]=\underbrace{(h_i^{D_r}[k])^H \Theta h_i^{D_t}[k]}_{ 
 Reflection\,link} +\underbrace{g_i^D[k]}_{Direct\,link}.
\end{array}
\end{equation}
In this equation, the direct link is
\begin{equation*}
\begin{array}{l}
g_i^D[k]= L(d_i^D) r_i^D[k],
\end{array}
\end{equation*}
where $d_i^D$ is three dimension distance between $D_i^t$ and $D_i^r$. $m_i^D[k]$ denotes Rayleigh fading for $i$'th D2D pair on $k$'th subchannel, assumed to follow a complex Gaussian distribution with zero mean and unit variance, i.e. $r_i^D[k] \sim \mathcal{C} \mathcal{N} (0,1), \forall i\in M,\forall k\in K$.
In the reflection link, the channels for the RIS departure and arrival links, respectively, can be expressed as
\begin{equation*}
\begin{array}{l}
(h_i^{D_r}[k])^H = L(d_i^{D_r}) e^ {-j2 \pi \frac{d_i^{D_r}}{\lambda [k]}} \mathbf{a}_{AoD}^H [k]
\end{array}
\end{equation*}
and
\begin{equation*}
\begin{array}{l}
h_i^{D_t}[k] = L(d_i^{D_t}) e^ {-j2 \pi \frac{d_i^{D_t}}{\lambda [k]}} \mathbf{a}_{AoA}[k],
\end{array}
\end{equation*}
where $d_i^{D_t}$ and $d_i^{D_r}$ denote the distance between RIS and $D_i^t$, between RIS and $D_i^r$, respectively. The array responses for the RIS upon arrival and departure are indicated as
\begin{equation*}
\begin{array}{l}
\mathbf{a}_{AoD}[k] = \big[ 1, ... , e^ {-j2 \pi \frac{d_i^{D_t}}{\lambda [k]} (N-1) sin(\theta _{AoD} )} \big]^T
\end{array}
\end{equation*}
and
\begin{equation*}
\begin{array}{l}
\mathbf{a}_{AoA}[k] = \big[ 1, ... , e^ {-j2 \pi \frac{d_i^{D_t}}{\lambda [k]} (N-1) sin(\theta _{AoA} )} \big]^T,
\end{array}
\end{equation*}
respectively. Here $\theta _{AoD}$ and $\theta _{AoA}$ are the departure and arrival angles of the RIS, respectively. The phase shift and amplitude attenuation $\\alpha$ for each RIS element can be written as $\Theta = diag [\alpha e^{j \theta_1} ,\alpha e^{j \theta_2} ,...,\alpha e^{j \theta N} ]$ where $ \alpha \in [0, 1]$ and $ \theta \in [0, 2 \pi)$. 
\begin{enumerate}
\item{CU-to-BS:}
The message signal transmitted by the $k$'th CU to the BS over the $k$'th RB is given as 
\begin{equation}
\begin{array}{l}
y_{k,BS}[k] =h_{k,BS}[k]\,x_k+h_{i,BS}[k]\,x_i+z,
\label{CU-to-BS}
\end{array}
\end{equation}
Where $x_i = \sqrt{p_i} u_i$ and $x_k = \sqrt{p_k} u_k$ denote the signal from $D_i^t$ and $CU_k$, respectively. \(p_i\) and \(p_k\) represent the transmit power of \(D_i^t\) and \(CU_k\), respectively. \(u_i\) and \(u_k\) are the entries with unit variance and zero mean, while \(z \sim \mathcal{N}(0, \sigma^2)\) denotes the AWGN noise signal with a mean of zero and variance \(\sigma^2\).
According to (\ref{CU-to-BS}), the SINR received at the BS over the $k$'th
RB is given as
\begin{equation}
\begin{array}{l}
\gamma_{k,BS}[k] = \frac{p_k\, \big|h_{k,BS}[k]\big|^2}{\sum_{i=1}^I \rho_{k,i}\, p_i\, \big|h_{i,BS}[k]\big|^2 +\sigma^2  },
\end{array}
\end{equation}
where $\rho_{k,i}$ is the RB allocation indicator. If $\rho_{k,i}=1$, the $k$'th CU and the $i$'th D2D pair reuse the same $k$'th RB, and zero otherwise.
\item{$D_i^t$-to-$D_i^r$:}
The message signal transmitted by the $D_i^t$ to the $D_i^r$ over the $k$'th RB is given as
\begin{equation}
\begin{array}{l}
y_i[k] =h_i[k]\,x_i + h_{k,i}[k]\,x_k +  h_{l,i}[k]\,x_l+z.
\end{array}
\label{$D_i^t$-to-$D_i^r$}
\end{equation}
According to (\ref{$D_i^t$-to-$D_i^r$}), the SINR received at the $D_i^r$ over the $k$'th RB is given as
\begin{equation}
\begin{array}{l}
\gamma_i[k] = \frac{p_i\, \big|h_i [k]\big|^2}{\rho_{k,i}\, p_k\, \big|h_{k,i} [k]\big|^2  + \sum_ {l=1  i\ne l}^ I \rho_{k,l}\, p_l \big|h_{l,i} [k]\big|^2 +\sigma^2}.   
\end{array}
\end{equation}
\item{CU-to-E:}
The message signal transmitted by the $k$'th CU to the $e$'th eavesdropper over the $k$'th RB is given as
\begin{equation}
\begin{array}{l}
y_{k,e}[k] =h_{k,e}[k]\,x_k + h_{i.e}[k]\,x_i+z.
\end{array}
\label{CU-to-E}
\end{equation}
According to (\ref{CU-to-E}), the SINR received at the $e$'th eavesdropper over the $k$'th RB is given as 
\begin{equation}
\begin{array}{l}
\gamma_{e}^{CU}[k] = \frac{p_k\, \big|h_{k,e} [k]\big|^2}{\sum_{i=1}^I \rho_{k,i}\, p_i\, \big|h_{i,e}[k]\big|^2 +\sigma^2  }.
\end{array}
\end{equation}
\item{$D_i^t$-to-E:}
The message signal transmitted by the $D_i^t$ to the $e$'th eavesdropper over the $k$'th RB is given as 
\begin{equation}
\begin{array}{l}
y_{i,e}[k] =h_{i.e}[k]\,x_i+ h_{k,e}[k]\,x_k + h_{l.e}[k]\,x_l+z.
\end{array}
\label{$D_i^t$-to-E}
\end{equation}
According to (\ref{$D_i^t$-to-E}), the SINR received at the $e$'th eavesdropper over the $k$'th RB is given as
\begin{equation}
\begin{array}{l}
\gamma_{e}^{DU}[k] = \frac{p_i\, \big|h_{i,e} [k]\big|^2}{p_k\, \big|h_{k,e} [k]\big|^2  + \sum_ {l=1  i\ne l}^ I \rho_{k,l}\, p_l \big|h_{l,e} [k]\big|^2 +\sigma^2}.   
\end{array}
\end{equation}
\end{enumerate}
According to Shannon's capacity theorem, the achievable rate of $i$'th D2D pair and $k$'th CU are, respectively, given by
\begin{equation} \label{Ri}
\begin{array}{l}
R_i=log_2(1+\gamma_i[k]),
\end{array}
\end{equation}
\begin{equation} \label{Rk}
\begin{array}{l}
R_k=log_2(1+\gamma_k[k]).
\end{array}
\end{equation}
And the achievable rate of $e$'th eavesdropper intercepting the information of $i$'th D2D pair and $k$'th CU, respectively, can be expressed as
\begin{equation} \label{Rie}
\begin{array}{l}
R_{i,e}=log_2(1+\gamma_{e}^{DU}[k]),
\end{array}
\end{equation}
\begin{equation} \label{Rke}
\begin{array}{l}
R_{k,e}=log_2(1+\gamma_{e}^{CU}[k]) .
\end{array}
\end{equation}
According to (\ref{Ri})-(\ref{Rke}) the secrecy ergodic capacity for the \(i\)-th D2D pair and the \(k\)'th CU over the \(k\)'th RB can be expressed as
\begin{equation} \label{SC_i}
\begin{array}{lll}
SC_i[k]=\mathbb{E} \Bigg[  B[k] \Big[ R_i - \underset{\substack{\mathbf{e \in (1,...,E)} }}{\operatorname{\text{max}}} & {R_{i,e}} \Big]^+ \Bigg]
\end{array}
\end{equation}
and\\
\begin{equation} \label{SC_k}
\begin{array}{lll}
SC_k[k]=\mathbb{E} \Bigg[  B[k] \Big[ R_k - \underset{\substack{\mathbf{e \in (1,...,E)} }}{\operatorname{\text{max}}} & {R_{k,e}} \Big]^+ \Bigg].
\end{array}
\end{equation}
The operator \( E[\cdot] \) represents the statistical expectation of \( [\cdot] \), indicating the expected rate over the small-scale fading distribution. Additionally, \( B[k] \) denotes the bandwidth of the \( k \)th subchannel. The SSC of the overall network using (\ref{SC_i}) and (\ref{SC_k}) is given as 
\begin{equation} \label{SC_{sum}}
\begin{array}{lll}
SC_{sum}=\sum_{k=1}^K \big( SC_k[k]+ \sum_{i=1}^M \rho_{k,i}\,  SC_i \big).
\end{array}
\end{equation}
The goal is to maximize equation (\ref{SC_{sum}}) by simultaneously optimizing the phase shifts of the RIS, the position of the RIS, the resource reuse coefficient $\mathbf{\rho} = [\rho_{1,1}, \dots, \rho_{1,M}, \dots, \rho_{K,1}, \dots, \rho_{K,M}]$, and the transmit power $\mathbf{p}^D = [p_1^D, \dots, p_M^D]$ of the D2D transmitters. This joint optimization problem is formulated as 
\begin{equation}\label{main problem}
\begin{aligned}
& \max_{\{\mathbf{L}^{RIS}, \Theta, \mathbf{\rho}, \mathbf{p}^D\}} SC_{\text{sum}} \\
\text{subject to: } & \, C_1:\quad p_i^D \leq p_{\text{max}}^D, \quad \forall i \in M \\
& C_2:\quad \gamma_i^D \geq \gamma_{\text{min}}^D, \quad \forall i \in M \\
& C_3:\quad \gamma_k^U \geq \gamma_{\text{min}}^U, \quad \forall k \in K \\
& C_4:\quad \rho_{k,i} \in \{0,1\}, \quad \forall i \in M, \forall k \in K \\
& C_5:\quad \sum_{k=1}^{K} \rho_{k,i} \leq 1, \quad \forall i \in M\\
& C_6:\quad 0 \leq \Theta_n \leq 2\pi,   \quad \forall n \in N\\
& C_7:\quad \mathbf{L}^{RIS} \in \mathbf{L}  , 
\end{aligned}
\end{equation}
where ($C_1$) ensures that the transmit power of each D2D transmitter does not surpass the predefined maximum power limit, which is essential for managing interference and ensuring efficient use of resources in the network. $\gamma_{min}^D$ and $\gamma_{min}^U$ represent the minimum required SINR at the D2D receiver and BS, respectively. Equations ($C_2$) and ($C_3$) ensure that the received signal power is sufficiently greater than the interference and noise levels at both the D2D receiver and the BS, thereby maintaining the quality of the communication links. Constraints ($C_4$) and ($C_5$) assume that each D2D pair is restricted to using a single RB, which makes the problem non-convex.
($C_6$) represents the limited phase shift interval of RIS due to hardware constraints, where only discrete phase shifts are considered.
The RIS cannot be installed everywhere, making continuous two-dimensional variables for its location impractical. Consequently, the location of the RIS is restricted to several discrete grid points, represented as $\mathbf{L} = \{\mathbf{L}_1, \mathbf{L}_2, \dots, \mathbf{L}_x\}$ where $x$ indicates the number of permissible grid points for RIS installation. The final constraint ($C_7$) reflects this limitation.
\section{Problem Solving}
To tackle this problem, we break it down into two subproblems. One subproblem aims to optimize resource allocation for D2D communications, while the other focuses on optimizing the positioning and phase shifts of the RIS. This decomposition not only significantly reduces the computational load on the BS but also allows the D2D pairs to determine their resource-sharing policies using local information, thus minimizing transmission overhead. In this section, we first introduce the fundamental concepts of reinforcement learning (RL) and then proceed to solve (\ref{main problem}).
\subsection{Mathematical Details}
RL is a dynamic process wherein an agent interacts with its environment to acquire knowledge and develop a behavior that aims to maximize cumulative rewards. At every discrete time step, denoted as $t$, the agent strategically selects actions, represented as $a^{(t)} \in A$, guided by its policy $\pi$: $S \rightarrow A$. This decision-making process results in the agent receiving a reward, denoted as $r^{(t)}$, and results in a transition to a new state within the environment, represented as $s^{(t)}$. The return is defined as the sum of discounted rewards, expressed as $R_t = \sum_{i=t}^T {\gamma^{i-t}r(s^{(i)}, a^{(i)})}$ where $\gamma$ serves as a discount factor that regulates the emphasis on short-term. After each time step $t$, the tuple $e^{(t)}=(s^{(t)}, a^{(t)}, r^{(t)}, s^{(t+1)})$ that is called transition, is stored in the replay buffer  $\mathcal{B}$ with size $\mathcal{D}$ for use in calculating the loss functions \cite{8917869}.
In RL, the objective is to identify the strategy \(\pi\) that maximizes the expected return. This is achieved by optimizing the action-value function, denoted as $q^{\pi}(s,a) = \mathbf{E_\pi} \big[ R_t | s^{(t)} = s, a^{(t)} = a \big]$ where the goal is to select the optimal policy \(\pi^*\) that maximizes the expected value. The optimal action-value function \(q^*\) is given by $q^*(s^{(t)}, a^{(t)}) = \rm{E} \big[ r^{(t)} + \gamma \underset {a^{'} \in A}{\text{max}} \{ q^*(s^{(t+1)}, a^{'}) | s^{(t)}, a^{(t)} \} \big]$.

To show that \(q(s^{(t)}, a^{(t)}) \to q^*(s^{(t)}, a^{(t)})\) as \(t \to \infty\), we can use the Bellman optimality equation for reinforcement learning. The approach relies on iterating over the state-action pairs until convergence to the optimal Q-values. Due to the impracticality of discrete training steps, neural networks, denoted as \(W\), are used to approximate the function that estimates the action-value function, i.e., $q(s^{(t)}, a^{(t)}) \approx q^*(s^{(t)}, a^{(t)})$,
which is the fundamental idea behind the Deep Q-Network (DQN). During training, a minibatch \((s^{(j)}, a^{(j)}, r^{(j)}, s^{(j+1)})\) is sampled from the training dataset. 

The goal of the training process is to minimize the error between the estimated Q-value and the true Q-value, which is formulated as $\text{Loss}(W) = \rm{E} \big[(q_{\text{target}} - q(s_j, a_j;W))^2\big]$,
where the target Q-value \(q_{\text{target}}\) is given by $q_{\text{target}} = r^j + \gamma \, q \big(s^{j+1}, \underset {a^{'} \in A}{\text{arg max}} \{q(s^{j+1}, a^{'}; W); W^-\}\big)$,
where \(q_{\text{target}}\) represents the output from the Q-target network in the conventional Double Deep Q-Network (DDQN) algorithm. \(W\) and \(W^-\) are the weights of the evaluation network and the target network, respectively \cite{9690178}.

\subsection{Optimization of Resource Allocation for D2D Pairs}
The goal of this subproblem is to optimize the power allocation and frequency channel assignment for D2D pairs. It is approached in a decentralized manner using a DDQN framework for each device. Each device makes local decisions, which alleviates the computational burden on the BS. The training algorithm for the decentralized DDQN models at the D2D pairs is outlined in Algorithm 1.\\
Taking into account the arbitrary location and phase shift of the RIS, the resource allocation optimization problem can be reduced to

\begin{equation}
\begin{aligned}
& \max_{\{\mathbf{\rho}, \mathbf{p}^D\}} SC_{\text{sum}} \\
\text{subject to: } & \quad C_1 - C_5.
\end{aligned}
\end{equation}
\begin{algorithm}
\caption{Decentralized DDQN-Based Secure Resource Allocation}
\begin{algorithmic}[1]
\STATE \textbf{Input:} Observation space $z_i$, SINR thresholds $\gamma_{\text{min}}^{D}$ and $\gamma_{\text{min}}^{U}$.
\STATE Initialize decentralized DDQN models $W_i$ for each D2D pair.
\FOR{each epoch}
    \STATE Randomly initialize RIS configuration and update channels.
    \FOR{each training step $t$}
        \FOR{each D2D pair $i$}
            \STATE Observe state $z_i^{(t)}$ and select action $a_i^{(t)}$ using $\epsilon$-greedy.
        \ENDFOR
        \STATE Compute secure reward
        \[
        r^{(t)} = 
        \begin{cases}
        \xi \, SC_{\text{sum}}, & \text{if constraints satisfied}, \\
        0, & \text{otherwise}.
        \end{cases}
        \]
        \STATE Store transitions and update decentralized DDQN models.
    \ENDFOR
\ENDFOR
\end{algorithmic}
\end{algorithm}
\subsection{RIS Positioning and Phase Shift Optimization}
The objective of this subproblem is to determine the optimal position of the RIS and adjust the phase shifts of its elements to enhance the overall performance of the network. The solution is computed centrally at the BS using a centralized DDQN algorithm, with the detailed steps outlined in Algorithm 2. This method effectively mitigates the overestimation issue commonly encountered in traditional DQN algorithms.\\
With this resource-sharing information, the capacity optimization problem at the BS can be expressed as
\begin{equation}
\begin{aligned}
& \max_{\{\mathbf{L}^{RIS}, \mathbf{\Theta} \}} SC_{\text{sum}} \\
\text{subject to: } & \quad C_1, C_2,C_3, C_6, C_7.
\end{aligned}
\end{equation}
\begin{algorithm}
\caption{Centralized DDQN-Based RIS Optimization}
\begin{algorithmic}[1]
\STATE \textbf{Input:} RIS position $\mathbf{L}^{RIS}$, phase shifts $\mathbf{\Theta}$.
\STATE Set up the action-value function $Q$ and the replay memory.
\FOR{each epoch}
    \STATE Update large-scale fading channels.
    \FOR{each training step $t$}
        \STATE Observe state $S^{(t)} = [\mathbf{\rho}, \mathbf{L}^{RIS}, \mathbf{\Theta}]$.
        \STATE Select action $a^{(t)}$ for RIS configuration using $\epsilon$-greedy.
        \STATE Compute secure reward
        \[
        r^{(t)} = 
        \begin{cases}
        \xi \, SC_{\text{sum}}, & \text{if constraints satisfied}, \\
        0, & \text{otherwise}.
        \end{cases}
        \]
        \STATE Store transitions and update centralized DDQN model.
    \ENDFOR
\ENDFOR
\end{algorithmic}
\end{algorithm}

\section{Simulation Results}
\begin{figure*}[!t]
\centering
\subfloat[]{\includegraphics[width=3.5in]{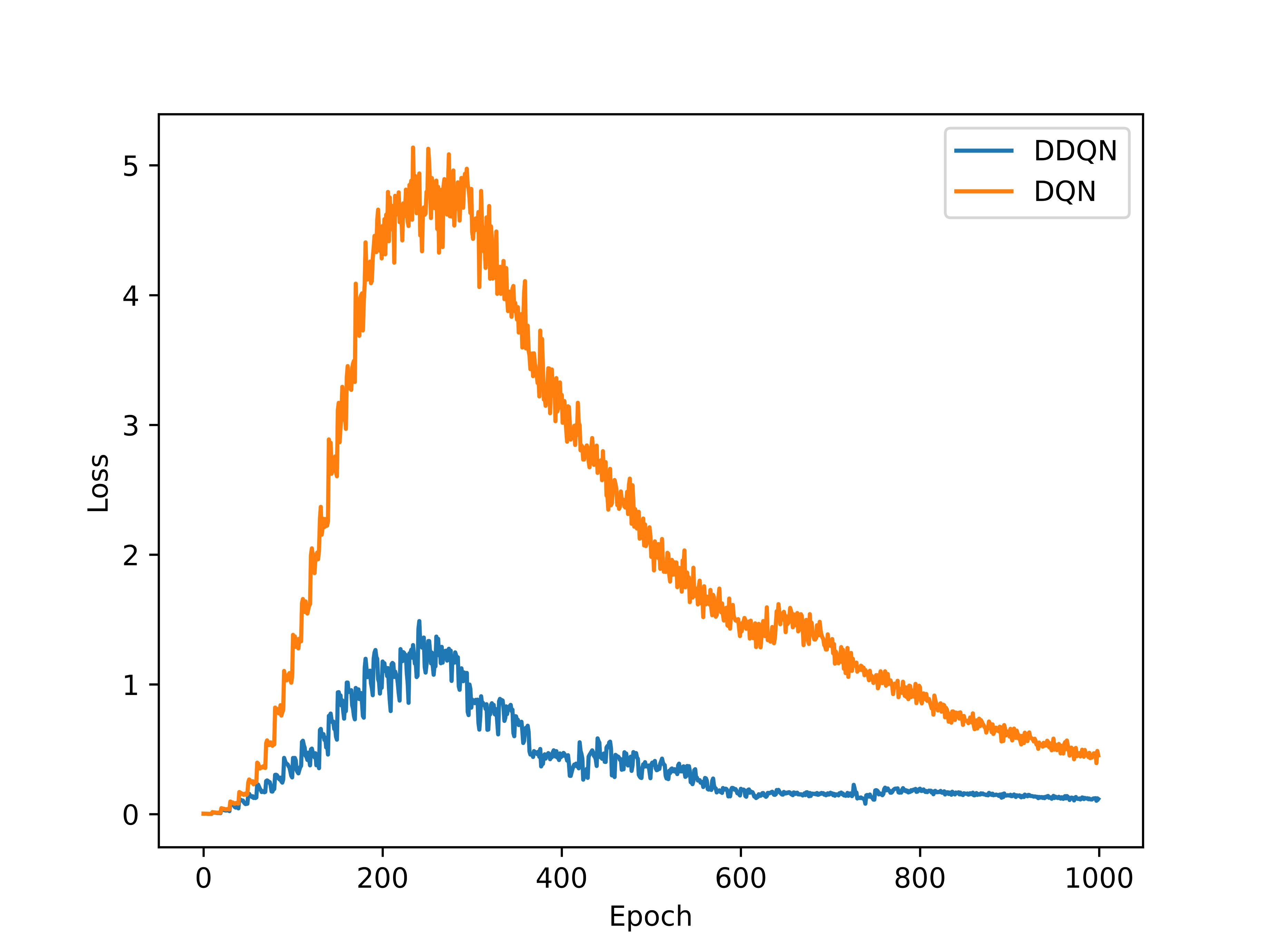}%
\label{LOSS}}
\hfil
\subfloat[]{\includegraphics[width=3.5in]{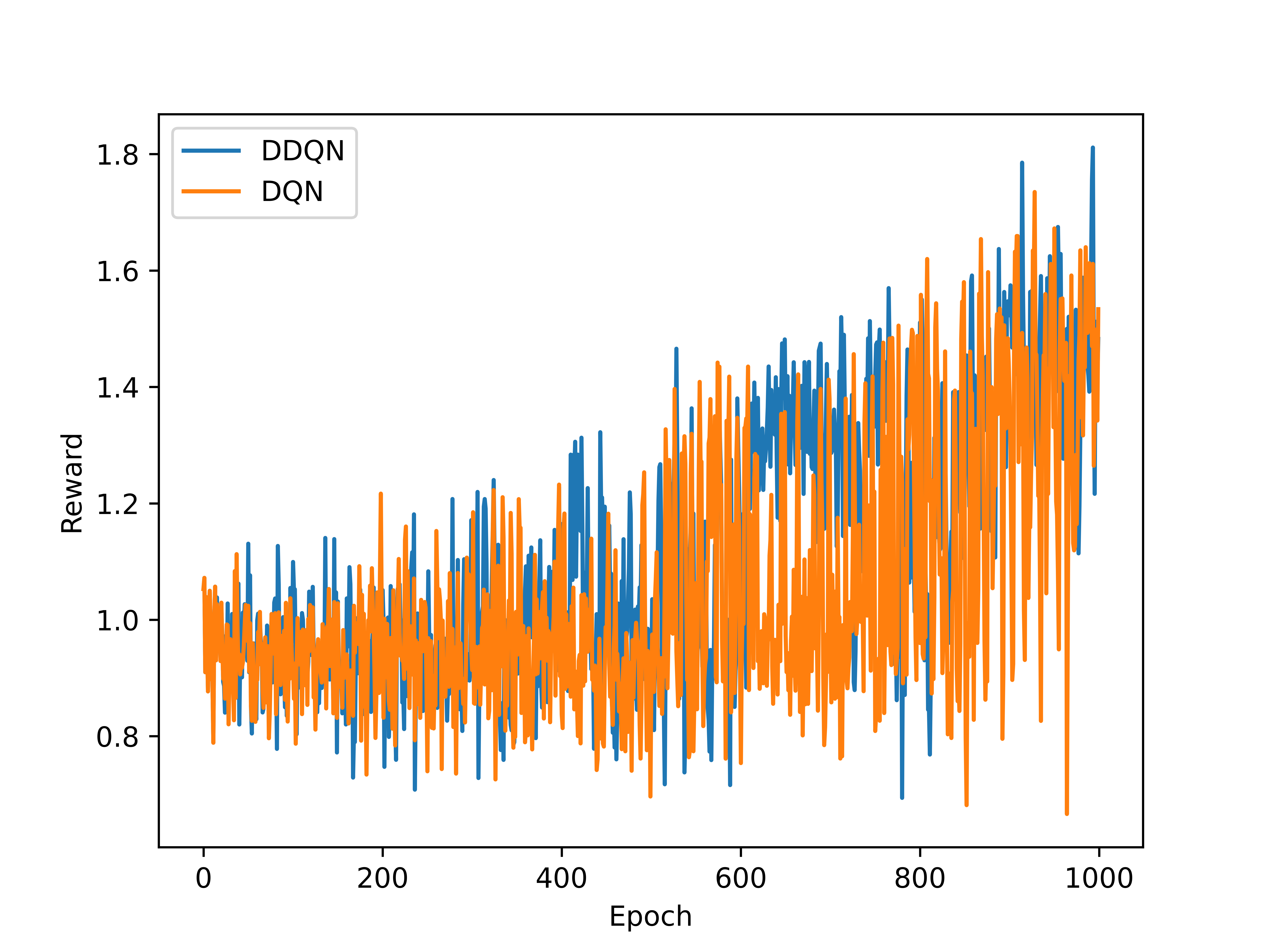}%
\label{REWARD}}
\caption{Evaluating training efficiency for RIS optimization at the BS using pre-trained decentralized framework for resource allocation.}
\label{fig_sim}
\end{figure*}
This section provides a comprehensive analysis of the performance, effectiveness, and robustness of the DDQN method within the proposed system model. The experimental setup consists of four D2D pairs, four cellular users, a BS, and a RIS equipped with $N = 8$ elements, all distributed within a 100m by 100m area. This area is divided into $x = 16$ equal sections to allow flexibility in the placement of the RIS. The antenna heights are set at 1.5 m for cellular users and D2D pairs, 25 m for BS, and 10 m for RIS, with the frequency of each subchannel fixed at 1 MHz. The transmission powers are 23 dB for the cellular mode and in the range of [0, 24] dB for D2D communications. \\
The decentralized DDQN architecture consists of five fully connected neural network layers, with three hidden layers containing 500, 250, and 120 neurons, respectively. Training is performed using the Rectified Linear Unit (ReLU) activation function and the RMSProp optimizer. The resource allocation model is updated only when significant changes occur in the wireless communication system, while it remains fixed during RIS optimization.\\
Figures \ref{LOSS} and \ref{REWARD} depict the neural network's loss and reward at the BS during RIS optimization, highlighting the effectiveness of the DDQN method. Specifically, Figure \ref{LOSS} demonstrates the superiority of DDQN compared to DQN, showing consistently lower loss values across all training epochs and faster convergence, which can be attributed to the mitigation of overestimation issues present in DQN. Moreover, Figure \ref{REWARD} reinforces these performance improvements, as DDQN achieves higher average rewards over successive training epochs compared to DQN.\\
\begin{figure}[!t]
\centering
\includegraphics[width=3.4in]{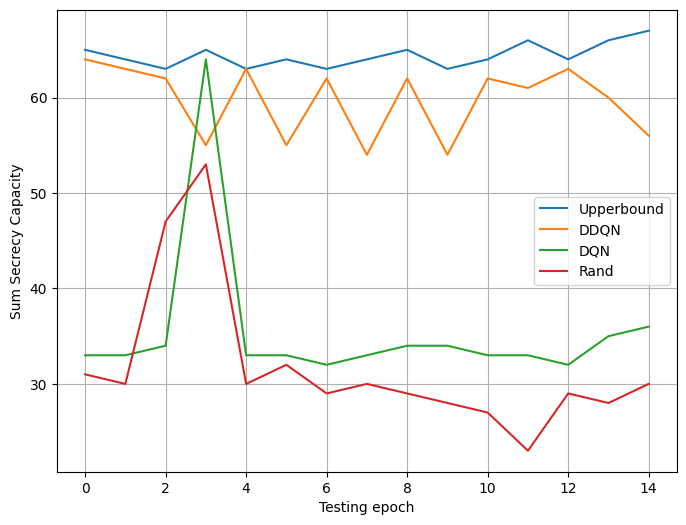}
\caption{Performance of secure resource allocation.}
\label{performance}
\end{figure}
\begin{figure}[!t]
\centering
\includegraphics[width=3.9in]{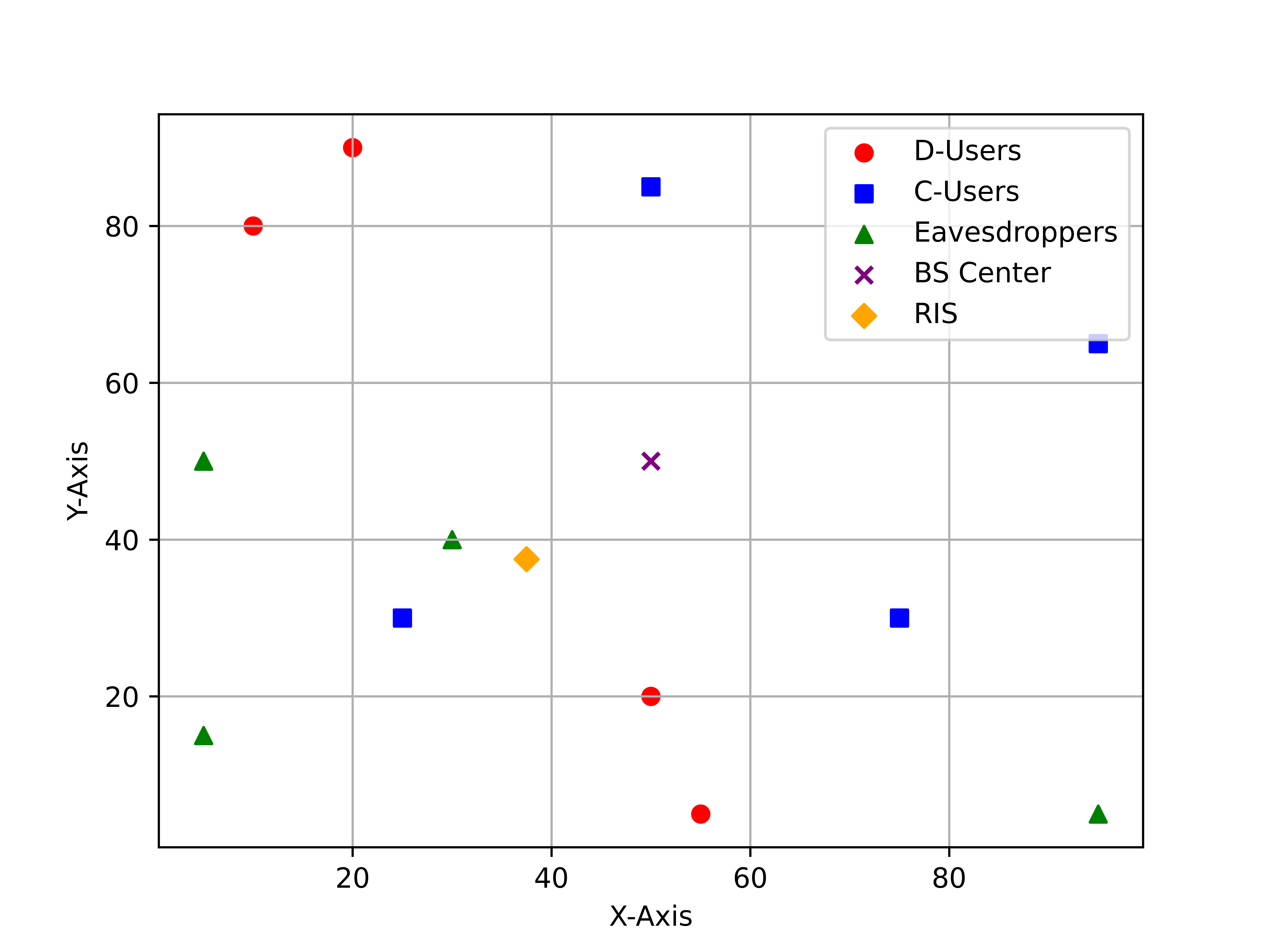}
\caption{Optimization of RIS Location.}
\label{location}
\end{figure}
Figure \ref{performance} illustrates the robustness of the decentralized DDQN method, tested in various random configurations of the RIS phase shifts and position in each testing epoch. Comparative evaluations with random allocation, decentralized DQN, decentralized DDQN, and exhaustive search (upper bound) demonstrate that the decentralized DDQN method achieves near-optimal performance, with faster convergence and reduced resource consumption. This makes it a favorable alternative to centralized DDQN, offering similar performance with less overhead. In contrast, the decentralized DQN method shows a higher SSC, highlighting its ineffectiveness in maximizing the objective function and satisfying the SINR requirements for both the BS and the D2D receivers.\\
Figure \ref{location} presents the performance of the centralized DDQN method, providing insight into the optimal location that yields the highest reward. The results indicate that the RIS is strategically placed to minimize the distance to any device within the network.\\
\section{Conclusion}
This paper introduces a comprehensive system model aimed at optimizing the secure performance of both D2D and cellular users in the presence of eavesdroppers, with the goal of maximizing the system's sum secrecy capacity. Due to the problem's complexity and nonconvexity, a decentralized double deep Q-network (DDQN) framework is used for resource allocation among D2D pairs, while a centralized DDQN framework is applied for the optimization of RIS location and phase shifts. To simplify the optimization process, the joint task is broken down into manageable subproblems, which alleviates the computational load on the central base station through decentralized resource allocation. Simulation results demonstrate the effectiveness of the proposed approach.

{\bibliographystyle{ieeetr}
\onehalfspacing
\bibliography{Arxiv_V}}

\vfill

\end{document}